\documentclass{INTERSPEECH2023}
\usepackage{subfigure}

\usepackage{wrapfig}

\title{The FruitShell French synthesis system at the Blizzard 2023 Challenge}
\name{XinQi$^{1,2}$, XiaopengWang$^{1,2}$, ZhiyongWang$^{1,2}$,WangLiu$^{1,2}$,MingmingDing$^3$,ShuchenShi$^4$}
\address{
  $^1$State Key Laboratory of Multimodal Artificial Intelligence Systems, CASIA, Beijng, China\\
  $^2$School of Artificial Intelligence, University of Chinese Academy of Sciences, Beijing, China \\
  $^3$Anhui Province Key Laboratory of Multimodal Cognitive Computation, School of Computer Science and Technology, Anhui University,  Hefei, China
  $^4$ Shanghai Polytechnic University
  }
\email{\{qixin221,wangxiaopeng22\}@mails.ucas.ac.cn,wangzhiyong2022@ia.ac.cn,
\\wang.liu@nlpr.ia.ac.cn,e21301171@stu.ahu.edu.cn,20221513084@stu.sspu.edu.cn}

\begin{document}

\maketitle

\begin{abstract}
This paper presents a French text-to-speech synthesis system for the Blizzard Challenge 2023. The challenge consists of two tasks: generating high-quality speech from female speakers and generating speech that closely resembles specific individuals. Regarding the competition data, we conducted a screening process to remove missing or erroneous text data. We organized all symbols except for phonemes and eliminated symbols that had no pronunciation or zero duration. Additionally, we added word boundary and start/end symbols to the text, which we have found to improve speech quality based on our previous experience. For the Spoke task, we performed data augmentation according to the competition rules. We used an open-source G2P model to transcribe the French texts into phonemes. As the G2P model uses the International Phonetic Alphabet (IPA), we applied the same transcription process to the provided competition data for standardization. However, due to compiler limitations in recognizing special symbols from the IPA chart, we followed the rules to convert all phonemes into the phonetic scheme used in the competition data. Finally, we resampled all competition audio to a uniform sampling rate of 16 kHz. We employed a VITS-based acoustic model with the hifigan vocoder. For the Spoke task, we trained a multi-speaker model and incorporated speaker information into the duration predictor, vocoder, and flow layers of the model. The evaluation results of our system showed a quality MOS score of 3.6 for the Hub task and 3.4 for the Spoke task, placing our system at an average level among all participating teams. 
\end{abstract}
\noindent\textbf{Index Terms}: Data cleaning, random duration prediction, speech synthesis

\section{Introduction}

The Blizzard Challenge is held once a year, and this year's theme is French synthesis. Participants are required to synthesize French text for two tasks within a limited time frame, using a limited dataset provided by the organizers. 

The two tasks are the Hub task and the Spoke task. The Hub task involves building a voice using the provided French data (NEB) using only publicly available data. The goal is to synthesize a voice that closely resembles the characteristics of the NEB dataset. On the other hand, the Spoke task requires participants to build a voice that is as close as possible to the AD dataset, using the provided French data. 

In recent years, text-to-speech synthesis systems have been a hot topic of research.  The goal of these systems is to take specific input text and generate high-quality, natural-sounding speech audio while trying to make the voice and prosody as close as possible to the desired target. 

Speech synthesis has undergone various evolutions in its technology. The earliest methods involved physical modeling of speech production, simulating the human vocal tract to synthesize speech \cite{1454423,shadle2001prospects}. This was followed by waveform concatenation methods for speech synthesis \cite{1170350,black1998festival}, where audio segments were selected from a database and concatenated to generate speech waveforms. However, this approach required large databases. 

Subsequently, statistical parametric synthesis methods were developed \cite{yoshimura1999simultaneous}. The basic idea was to generate acoustic-related parameters and then recover waveforms from these parameters. 

Over time, researchers have gradually developed methods for speech synthesis using deep neural networks. Initially, partial end-to-end synthesis was performed, where an acoustic model \cite{wang2017tacotron,ren2019fastspeech,li2019neural} generated acoustic features, which were then converted back to waveforms using a vocoder \cite{oord2016wavenet,valin2019lpcnet,kong2020hifi}. However, this approach can accumulate errors between different components, leading to a decrease in synthesis quality. 

To address this issue, researchers have proposed some end-to-end models such as Deep Voice 3 \cite{ping2017deep}, Fastspeech2 \cite{ren2020fastspeech}, GlowTTS \cite{kim2020glow}, and Diffspeech \cite{jeong2021diff}. These models can directly generate the final speech from input text. The aim is to simplify the text analysis module by directly using character/phoneme sequences as input and simplifying acoustic features into magnitude spectra to improve the accuracy of acoustic feature prediction. 

This article will provide a brief overview of an end-to-end approach used for French synthesis in the competition, including the complete process, including data preprocessing. The results of the competition will also be presented at the end.

\section{Methods}

\subsection{Overview}

\begin{figure}[h]
  \centering
  \includegraphics[width=\linewidth]{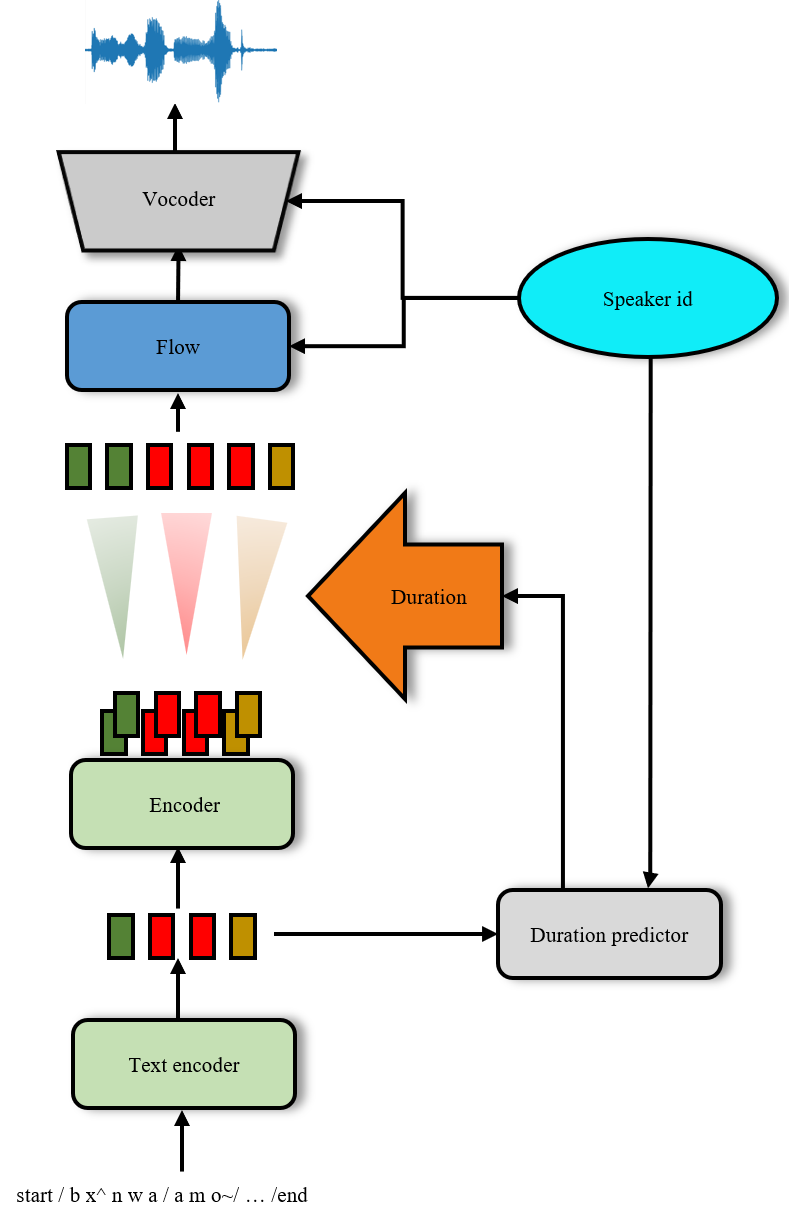}
  \caption{An overview of our system. }
  \label{An-overview-of-our-system}
\end{figure}

We adopted a fully end-to-end approach for French synthesis, with the VITS (Vocoder Inverse Text-to-Speech) model as the main framework\cite{kim2021conditional}. The main parameter count of the model is similar to VITS, approximately around 28 million. We simultaneously trained two models for two tasks, both utilizing 4 NVIDIA GeForce RTX 3090 graphics cards. For the Spoke task, we trained for approximately two million steps, which took about two weeks to complete. On the other hand, for the Hub task, we trained for approximately 500,000 steps, which took approximately 5 days. We initially set the training parameters as follows: 20,000 training epochs, a learning rate of 2e-4, and a batch size of 16. However, we did not complete the full 20,000 epochs. For the Hub task, we trained for 700 epochs, and for the Spoke task, we trained for 10,000 epochs. The overall architecture of the model consists of two components: an acoustic model and a vocoder, with an additional duration predictor. The vocoder was trained from scratch, meaning it was trained anew rather than using pre-existing weights or models.  Additionally, we incorporated speaker information to differentiate between speakers. The overall structure of the model is illustrated in Figure \ref{An-overview-of-our-system}. 

The task of the acoustic model is to convert the input text into acoustic features, which represent the spectrogram of the speech. In the VITS model, the acoustic model employs the structure of a Variational Autoencoder \cite{kingma2013auto}. 

The VAE consists of two main components: an encoder and a decoder. The encoder transforms the input text into a low-dimensional representation in the latent space, often referred to as a "latent vector" or "latent variable. " The decoder then maps the latent vector back to the representation of acoustic features. 

In addition to the encoder and decoder, the VITS model incorporates the concept of variational inference for training and generating acoustic features. Variational inference introduces a supplementary loss function, known as the Kullback-Leibler (KL) divergence, to ensure the proximity between the distribution of the latent vector and a predefined prior distribution. This helps generate smoother and more continuous acoustic features, thereby enhancing the naturalness of the synthesized speech. 

Our system utilizes the HiFiGAN vocoder. HiFiGAN is based on a generative adversarial network (GAN) architecture, consisting of a generator and a discriminator. The generator takes acoustic features as input and generates waveform samples that resemble natural speech. The discriminator is designed to distinguish between generated speech and real speech samples. The generator and discriminator are trained together through adversarial training, where the generator aims to generate speech that can deceive the discriminator, while the discriminator learns to accurately differentiate between real and generated speech. 

The unique aspect of HiFiGAN lies in its ability to generate high-fidelity and high-quality speech waveforms. It utilizes multi-scale discriminators and feature matching loss to ensure that the generated speech closely matches the target waveform in both spectral and temporal features. This results in synthesized speech that sounds more natural and closer to human speech. 

The random duration predictor in the VITS model is a special type of duration predictor that is used to introduce randomness and diversity into the speech synthesis process. Unlike traditional deterministic duration predictors, the random duration predictor generates different durations for each synthesized speech, making the generated speech more varied and natural. 

During the synthesis process, the output of the random duration predictor is used as the sampling distribution for durations. Each time speech is synthesized, a different duration is randomly sampled from this distribution and applied to the corresponding text unit. As a result, even with the same input text, slightly different speech outputs are generated each time. 

The introduction of the random duration predictor allows the VITS model to generate diversity and more natural speech. It increases the variation and expressiveness in speech synthesis, making the synthesized speech closer to the way human speech is performed. 

\subsection{Front-end processing}
\begin{figure}[h]
  \centering
  \includegraphics[width=\linewidth]{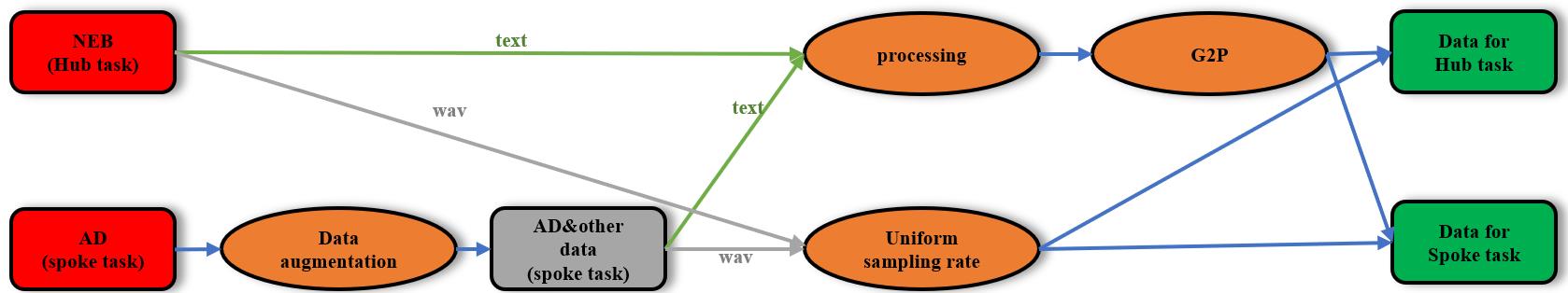}
  \caption{Front-end processing. }
  \label{Front-end-processing}
\end{figure}

The organizers provided two datasets for the competition: the NEB dataset, which consists of 289 chapters from 5 audiobooks from Librivox (51:12) read by Nadine Eckert-Boulet, and the AD dataset, which includes 2515 utterances (2:03) read by another female French speaker, Aurélie Derbier. Although these datasets provide corresponding text, they contain numerous errors, requiring data preprocessing. For the NEB dataset, we performed data splitting based on textual cues and retained a total of 61,330 audio samples. As for the AD dataset, after splitting and incorporating additional multi-speaker data, we conducted cleaning and ended up with a total of 15,000 retained samples. We divided the dataset into training and testing sets using an 80-20 split ratio, with 80\% of the data used for training and 20\% for testing. The complete preprocessing process can be seen in Figure \ref{Front-end-processing}. 

First, there were missing corresponding texts for some audio segments. Since we were unsure if our ASR method was consistent with the organizers', using our own transcription results could lead to a mismatch between the text and acoustic features. Therefore, we decided to discard the data with missing text. Secondly, there were transcription errors and misalignments in some data, so we also discarded these suspicious data. For the provided audio segments, which were all complete recordings, we split the audio into corresponding text segments based on the provided start and end time information. Finally, we obtained a relatively clean dataset. 

However, at this point, we couldn't directly use the data as the text still contained various punctuation marks, special symbols, and word boundary markers. According to the provided duration information, a significant portion of these symbols was not pronounced.  Hence, we filtered out these non-phonetic symbols and removed them from the text data, making the data more concise and efficient for training. We retained symbols with noticeable duration or those occurring at sentence pauses. Additionally, we added a period at the end of each sentence to indicate the completion of a full sentence. 

Regarding word boundary information, we kept it intact. Adding word boundary information in the text has been proven to enhance the quality of speech synthesis, allowing for reasonable adjustments in speech rate. However, the original data used curly braces to enclose each word, which was inconvenient for reading the text. Therefore, we replaced the curly braces with a slash "/" inserted at the word boundary positions, while keeping the remaining phonetic-level spacing represented by spaces. 

After that, we added additional start and end tags at the beginning and end of each sentence, respectively, to control the silence before and after the speech synthesis process.

\subsubsection{Data augmentation}
For the Spoke task, as there was limited specific-speaker data provided, it was challenging to train a good model directly. Therefore, we searched for high-quality open-source multi-speaker French datasets \cite{CommonVoice} and combined them with the provided NEB dataset to train a multi-speaker model. The additional dataset includes 75 hours of audio and over 1000 different speakers. This approach of augmenting the data aimed to enhance the quality of specific-speaker speech synthesis. The same data processing steps mentioned earlier were applied to these additional datasets. 

\subsubsection{G2P}
Since the provided data does not use the International Phonetic Alphabet (IPA), and most existing Grapheme-to-Phoneme (G2P) models rely on IPA, we did not use the phonetic transcription provided by the competition organizers. As our team had no prior experience with French, we opted to use open-source and widely recognized G2P models, such as Phonemizer\cite{bernard2021phonemizer}. We transcribed all the cleaned texts into IPA symbols and annotated them based on the corresponding word boundaries. However, this introduced another issue: the IPA symbols contain special characters that the compiler cannot recognize effectively. To address this, we replaced the IPA symbols with the phonetic transcription scheme provided in the competition data. 

\subsubsection{French Chanting}

In French, there is a special pronunciation rule called 'liaison,' where the final sound of one word carries over to the next word. In our text, we have specifically addressed this rule by processing according to the G2P results. We take the last phoneme of the preceding word and use it as the beginning of the next word, while replacing the space with a special symbol. 

\subsubsection{Uniform sampling rate}
The provided data had a sampling rate of 22.05 kHz, while the augmented data we found had a sampling rate of 16 kHz. To facilitate model training, we used SoX to downsample all the audio to a uniform sampling rate of 16 kHz. 

\section{Results}
In this section, we will primarily showcase the evaluation results of our system in the competition. We will present the results separately for each task. Our system is identified by the ID "S". 

\subsection{Hub task Results}
For the Hub task, the rules state that we can only use the NEB dataset provided by the organizers and cannot utilize any additional data. In this task, we will primarily showcase three metrics: quality, similarity, and pronunciation error rate. 

\subsubsection{Pronunciation error rate}
\begin{figure}[h]
  \includegraphics[width=\linewidth]{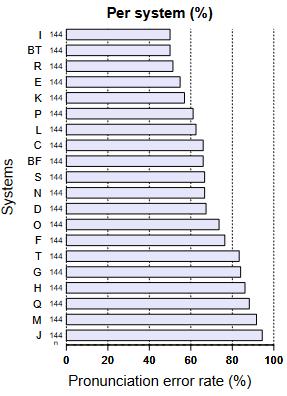}
  \caption{Pronunciation error rate. }
  \label{error-rate}
\end{figure}

In this section, we primarily present the results of our system's pronunciation error rate. As shown in Figure \ref{error-rate}, our system is ranked in the middle among all participating teams. 

\begin{figure*}[t]
  \centering
  \includegraphics[width=\linewidth]{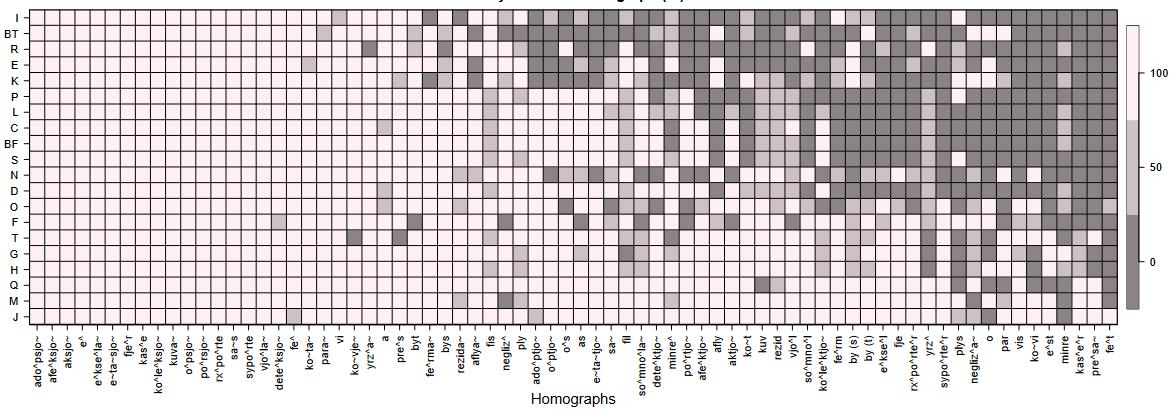}
  \caption{Per system and homograph. }
  \label{Homographs}
\end{figure*}

Additionally, we provide a distribution chart of the pronunciation error rate for the given homographs, as depicted in Figure \ref{Homographs}. Lighter cells indicate higher accuracy, and a greater number of light cells suggests better performance of the current system. 

\subsubsection{Quality}
\begin{figure}[h]
  \centering
  \includegraphics[width=\linewidth]{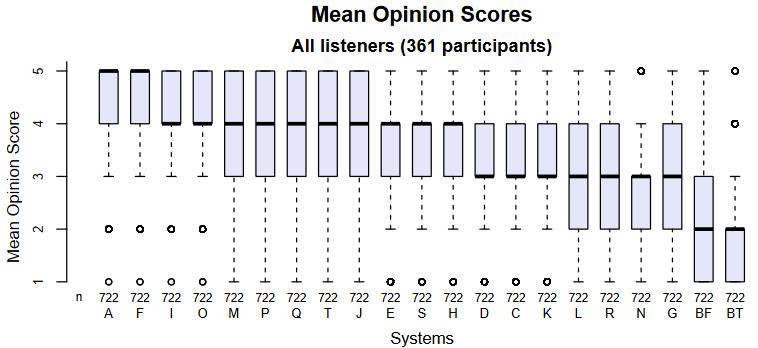}
  \caption{Hub task MOSquality. }
  \label{quality}
\end{figure}

This section presents the MOS (Mean Opinion Score) ratings of our system's generated speech quality, as shown in Figure \ref{quality}. Our specific MOS score has a median value of 4, which falls in the middle range compared to all the participating teams. 

\subsubsection{Similarity}
\begin{figure}[h]
  \centering
  \includegraphics[width=\linewidth]{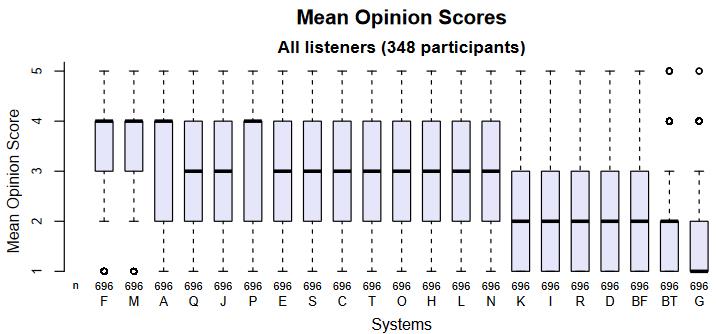}
  \caption{Hub task MOSsimilarity. }
  \label{Similarity}
\end{figure}

In the final section, we present the similarity scores of our system in the Hub task. As shown in Figure \ref{Similarity}, our similarity score is positioned in the upper-middle range, with a relatively small gap compared to the top-performing systems. 

\subsection{Spoke task Results}
This section primarily showcases the performance of our system in the Spoke task. Since this task has fewer evaluation metrics, we mainly focus on two indicators: quality and similarity. Compared to the Hub task, our scores may be slightly lower. We analyzed that the overall decline in audio quality for specific individuals is due to the lower quality of the additional data we found compared to the competition data. 

\subsubsection{Quality}
\begin{figure}[h]
  \centering
  \includegraphics[width=\linewidth]{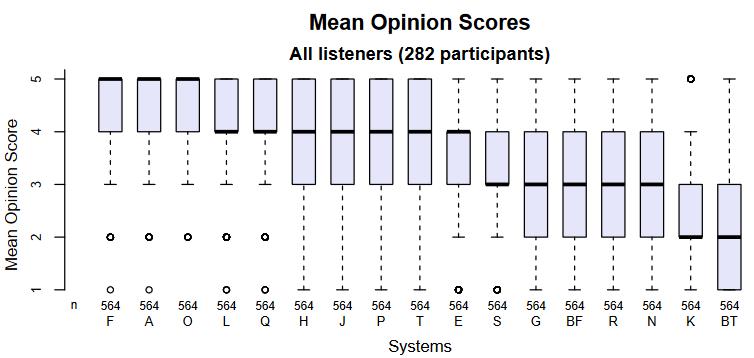}
  \caption{Spoke task MOSquality. }
  \label{AD_quality}
\end{figure}

This section showcases the MOS (Mean Opinion Score) ratings of the speech quality generated by our system, as depicted in Figure \ref{AD_quality}. Our system achieved an MOS score of 3.4 with a median of 3.6. 

\subsubsection{Similarity}
\begin{figure}[h]
  \centering
  \includegraphics[width=\linewidth]{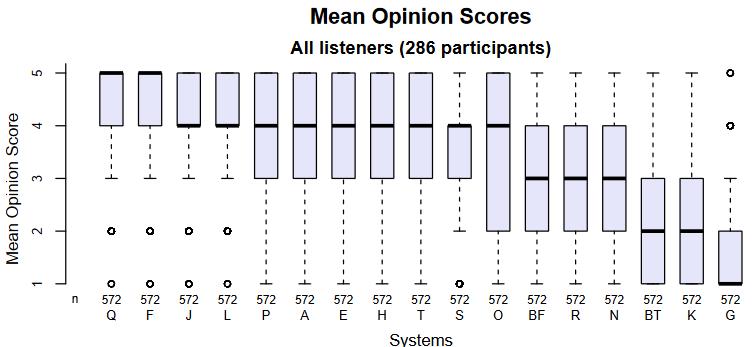}
  \caption{Spoke task MOSsimilarity. }
  \label{AD_Similarity}
\end{figure}

Our system achieved a similarity score of 3.5, as shown in Figure \ref{AD_Similarity}, indicating a moderate level of performance. 

\section{Conclusion}
In this paper, we presented a French text-to-speech synthesis system and provided a comprehensive overview of the entire process, from data preprocessing to speech synthesis. We thoroughly cleaned the data and annotated it according to the requirements of acoustic features. Additionally, we augmented the data based on the task requirements. The application of variational inference and a random duration predictor enhanced the realism of the synthesized audio. We utilized the state-of-the-art hifigan vocoder to generate high-quality audio. Finally, we showcased the evaluation results of the competition, where we achieved good performance across all the metrics. 

\section{Acknowledgments}

This work is supported by the National Natural Science Foundation of China(NSFC)
(No.62101553, No.61831022, No.U21B2010, No.61971419,
No.62006223, No.62276259, No.62201572, No.62206278). 

\bibliographystyle{IEEEtran}
\bibliography{mybib}

\end{document}